# Efficient light upconversion via resonant exciton-exciton annihilation of dark excitons in few-layer transition metal dichalcogenides

Yi-Hsun Chen[1,2]*[†], Ping-Yuan Lo[3][†], Kyle W. Boschen[2,4], Chih-En Hsu[5], Yung-Ning Hsu[5], Luke N. Holtzman[6], Guan-Hao Peng[3], Chun-Jui Huang[3], Madisen Holbrook[7], Wei-Hua Wang[8], Katayun Barmak[6], James Hone[9], Pawel Hawrylak[10], Hung-Chung Hsueh[5], Jeffrey A. Davis[2,4], Shun-Jen Cheng[3]*, Michael S. Fuhrer[1,2,11], Shao-Yu Chen[2,12,13]*

**AFFILIATIONS**

[1] School of Physics and Astronomy, Monash University, Clayton, Victoria 3800, Australia

[2] Australian Research Council Centre of Excellence in Future Low-Energy Electronics Technologies (FLEET), Monash University, Clayton, Victoria 3800, Australia

[3] Department of Electrophysics, National Yang Ming Chiao Tung University, Hsinchu 300 093, Taiwan.

[4] Optical Sciences Centre, Swinburne University of Technology, Hawthorn, VIC 3122, Australia.

[5] Department of Physics, Tamkang University, New Taipei City, 251301 Taiwan

[6] Department of Applied Physics and Applied Mathematics, Columbia University, New York, New York 10027, United States

[7] Department of Physics, Columbia University, New York NY 10027, USA

[8] Institute of Atomic and Molecular Sciences, Academia Sinica, Taipei 10617, Taiwan

[9] Department of Mechanical Engineering, Columbia University, New York, New York 10027, United States

[10] Department of Physics, University of Ottawa, Ottawa, Ontario, K1N 6N5, Canada

[11] Monash Centre for Atomically Thin Materials, Monash University, Clayton, 3800, VIC, Australia

[12] Center for Condensed Matter Sciences, National Taiwan University, Taipei 10617, Taiwan

[13] Center of Atomic Initiative for New Material, National Taiwan University, Taipei 10617, Taiwan

[†] Equally contributed

**ABSTRACT**

Materials capable of light upconversion—transforming low-energy photons into higher-energy ones—are pivotal in advancing optoelectronics, energy solutions, and photocatalysis. However, the discovery in various materials pays little attention on few-layer transition metal dichalcogenides, primarily due to their indirect bandgaps and weaker light-matter interactions. Here, we report a pronounced light upconversion in few-layer transition metal dichalcogenides through upconversion photoluminescence spectroscopy. Our joint theory-experiment study attributes the upconversion photoluminescence to a resonant exciton-exciton annihilation involving a pair of dark excitons with opposite momenta, followed by the spontaneous emission of upconverted bright excitons, which can have a high upconversion efficiency. Additionally, the upconversion photoluminescence is generic in $MoS_2$, $MoSe_2$, $WS_2$, and $WSe_2$, showing a high tuneability from green to ultraviolet light (2.34–3.1 eV). The findings pave the way for further exploration of light upconversion regarding fundamental properties and device applications in two-dimensional semiconductors.

**INTRODUCTION**

Upconversion photoluminescence (UPL) is an anti-Stokes phenomenon of light-matter interactions in which a material radiatively emits photons at an energy higher than the excitation energy. Since the process is able to generate high-energy photons, UPL is of interest in a wide range of applications across various fields such as biology[1–3], medicine[4,5], and energy[6–8]. Starting from the very first rare-earth doped nanomaterials[9], the demonstration of UPL has been reported in inorganic[10,11], organic[12,13], and organic-inorganic hybrid[14,15] semiconductors. Recent advancements, particularly in molecular systems employing triplet−triplet annihilation, have achieved high-quantum efficiency and/or low excitation density in the upconversion process.[16,17] However, there is still a growing demand for solid-state alternatives due to their enhanced durability and compatibility with existing semiconductor manufacturing technologies, which could open doors for broader applications.

Exciton-exciton annihilation (EEA) has a fundamental advantage over other upconversion mechanisms for spontaneous light upconversion in solid-state materials. EEA is a two-body non-radiative process in which one exciton transfers its energy and momentum to another, effectively doubling the energy of the resulting exciton.[18] The rate of EEA exhibits a quadratic dependence on exciton density[19], which is more efficient than the cubic dependence in three-body processes, such as Auger recombination[20], especially in the low-density regime. Despite its potential, EEA phenomena necessitate a dominance of excitonic effects in materials, a condition challenging to achieve in bulk semiconductors due to their small exciton binding energies. Atomically thin 2D semiconducting materials, on the other hand, possess excitons with binding energies reaching several hundreds of meV because of the reduced dielectric screening.[21] Moreover, they host a rich exciton landscape from the intriguing electronic band structures within the Brillouin zone, making them ideal candidates for exploring EEA-driven light upconversion. Indeed, non-radiative EEA has been demonstrated in low-dimensional materials such as carbon nanotubes[19,22], graphene nanoribbons[23], quantum wells[24], and quantum dots[25], and even 2D semiconductors[26–29]. In 2D semiconducting materials, the non-radiative Auger process in transition metal dichalcogenides (TMDs) has been intensively studied.[30–33] Despite the recent demonstration of light upconversion in TMDs[34–36], the UPL under the EEA principle has rarely been reported.

In this work, we report light upconversion in few-layer transition metal dichalcogenides (TMDs) via a resonant EEA of dark excitons. The EEA of dark excitons involves two momentum-indirect dark excitons with opposite momenta upconverting to higher-energy zero-momentum excitons at Γ valley, supported by layer-dependent dark/bright exciton energies observed in both experiments and theoretical calculations. The quadratic power dependence is characterised in $WSe_2$ atomic layers by using power dependent measurements, indicating EEA as the mechanism for the upconversion process. The resonance effect describes an energy overlap between upconverted excitons and high-energy bright exciton states, allowing efficient emission of upconverted excitons above the excitation energy. We also show that the UPL, initiated by using the continuous-wave (CW) laser with a fairly low-power density, is fundamentally different from second harmonic generation (SHG), which often necessitates high-power density through a pulsed laser. Lastly, our study showcases that UPL is remarkably tuneable through material choice ($WSe_2$, $WS_2$, $MoS_2$, and $MoSe_2$) and layer number, with upconversion photon energy ranging from 2.34 to 3.1 eV. Such tunability underscores the strong layer-dependent energy of dark excitons, originating from significant interlayer Coulomb interactions and quantum confinement effects inherent in atomically thin TMDs.

**RESULTS**

**Excitonic species and UPL in few-layer $WSe_2$.**

Figures 1a,b compare the absorbance and UPL spectra, respectively, of a 5L $WSe_2$ sample at $T$ = 80 K. The absorbance spectrum (Figure 1a) reveals four pronounced peaks attributed to bright excitons, which are labelled according to their electron/hole configurations, denoted as $MN_i$, where M and N represent the momentum of the electron and hole in the hexagonal Brillouin zone, respectively, and $i$ indicates the index in ascending energy order for identical configurations. The corresponding electron/hole configurations are schematically illustrated in Figure 1c. From low to high energy, the first and second peaks, labelled $KK_A$ and $KK_B$, correspond to the traditionally known A and B excitons respectively. At higher energy, based on the prior experiments[37] and the exciton spectra solved from the Wannier tight-binding (WTB)-based Bethe-Salpeter equation (BSE) incorporate with the quasiparticle band structures obtained from density functional theory with the $G_0W_0$ correction, we assign the peaks to QQ and ΓΓ excitons (further details in Methods). Notably, ΓΓ excitons are composed of electrons with negative effective mass and holes with positive effective mass. Such atypical conduction and valence bands may result in a nesting of excitonic bands with a high joint density of states.[38–40]

In the UPL spectroscopy, we collect luminescence signals spanning energies both below and above the excitation energy (2.33 eV), yielding a typical PL (upconversion PL) spectrum represented by the blue (ruby) curves in Figure 1b. Sharp features adjacent to 2.33 eV are attributed to Raman scattering by $WSe_2$. The PL emission from $KK_A$ and $KK_B$ excitons becomes exceedingly weak in 5L $WSe_2$ due to its indirect bandgap. Meanwhile, $KK_A$ emission is overwhelmed by the PL signals from $Al_2O_3$ substrate noted in grey shadow (see Supplementary Figure 1 for PL spectrum of 5L $WSe_2$ on a $SiO_2$/Si substrate). Below the $KK_A$ emission, we detect a faint PL peak from the lowest-energy momentum-indirect (dark) excitons, labelled as $X_D$, at approximately 1.31 eV, which is attributed to QK or QΓ excitons.[41,42] In the upconversion PL spectrum, we observe a notable emission of UPL around 2.64 eV, which is approximately twice the energy of the $X_D$ peak. Intriguing, the UPL intensity is significantly higher than the $X_D$ in the typical PL spectrum, suggesting an efficient light upconversion process. A detailed examination of the UPL emission (see the inset in Fig. 1b) reveals that the line profile can be modelled with two peaks: one at higher energy (coloured in orange) well-described by a Gaussian function and the other at lower energy (coloured in light green) described by a Lorentzian function (see Supplementary note 2). The Lorentzian profile suggests an ultrafast recombination lifetime.[43] The Gaussian peak primarily contributes to the UPL and is strongly correlated to $X_D$, which will be analysed in Figure 2; hence, we named it as upconverted excitons ($X_{up}$). The Lorentzian peak is located close to but not directly at the ΓΓ exciton energy seen in the absorbance spectrum. To be clear, we named it as high-lying excitons (HX) instead. The HX, which is different from contemporary studies[44], is defined as bright exciton states with photon energies higher than $KK_A$ and $KK_B$ excitons.

**Layer-dependent UPL and dark exciton characteristics.**

Figure 2a presents $X_D$ emissions of 2–6 L WSe$_2$, exhibiting a red shift in emission energy from 1.553 (2L) to 1.294 eV (6L), consistent with the evolution of the electronic band structure as the layer thickness increases.[45] Figure 2b shows the corresponding UPL spectra fitted with two components, HX and $X_{up}$, as mentioned in the inset of Figure 1b. The $X_{up}$ emission is located at 2.857 (3L), 2.712 (4L), 2.642 (5L), and 2.608 eV (6L), and the HX is located at 2.76 (3L) and 2.67 eV (4L) and 2.621 (5L). However, neither $X_{up}$ nor HX emission is detected in 2L WSe$_2$. For clarity, we summarise the energies of $X_{up}$, $X_D$ and ΓΓ excitons as a function of the number of layers in Fig. 2c. The ΓΓ exciton energies are extracted from absorbance spectra. The $X_{up}$ energies (filled blue circles) closely match the red dashed curve, which depicts twice the energy of $X_D$ (labelled as 2×$E_{XD}$), reinforcing that $X_{up}$ is strongly correlated with $X_D$. Furthermore, the $X_{up}$ energy moves toward ΓΓ excitons (filled green triangles) with the increasing number of layers and overlaps ΓΓ excitons beyond 5L WSe$_2$, resulting in a strongly enhanced light upconversion intensity evident in Figure 2b and quantified by the open blue triangles in Figure 2c. We found that upconversion efficiency $Q_{up}$ (defined in Methods), is greater than unity in 5L WSe$_2$ and above, which is at least ten times higher than 3L WSe$_2$. The $Q_{up} > 1$ means that UPL emission is larger than $X_D$ emission, quantitatively suggesting an efficient light upconversion occurred at the resonance between $X_{up}$ and ΓΓ excitons.

**Theoretical calculations of resonant exciton-exciton annihilation in few-layer WSe$_2$**

Figure 3a illustrates a resonant EEA process responsible for the efficient light conversion. Upon optical excitation, photoexcited bright excitons thermally relax to X$_D$ via multiple phonon scattering. X$_D$, momentum-forbidden dark excitons with a finite centre of mass (CoM) momentum, which population lifetime up to tens of nanoseconds[46,47] is orders of magnitude longer lifetime than bright excitons and thus favours pronounced many-body interactions[48,49], is upconverted to a higher energy state X$_{up}$ through the resonant EEA, followed by radiative recombination at an energy above the initial excitation.

We propose a model wherein a resonant EEA facilitates the light upconversion in few-layer WSe$_2$ based on the many-body theory for excitons, which microscopically describes the underlying electron and/or hole scattering processes in the EEA (see more detail in Supplementary note 3). The resonant EEA process involves the initial state (a pair of low-lying indirect dark excitons) upconverting to the final state (a high-lying bright exciton) under a fundamental restriction—the conservation of energy and momentum. Figure 3b illustrates a viable pathway for QΓ dark excitons undergoing the resonant EEA upon the momentum-conserving principle. In the right panel, the final state entails that both electron and hole are at the Γ valley, where bright excitons exhibit zero CoM momentum. For this to occur, two X$_D$ in the initial state (left panel) could be momentum-indirect with opposite momenta ($\mathbf{Q}_1$ and $\mathbf{Q}_2 = -\mathbf{Q}_1$). For example, considering a pair of QΓ/ Q'Γ excitons, where the hole locates at Γ valley, and the electron occupies the Q/Q' valley. The EEA of such dark exciton pairs describes that one dark exciton (X$_{D1}$ for example) recombines non-radiatively and transfers its energy and momentum to the other dark exciton (X$_{D2}$) in non-equivalent valley, and that the electron of the other dark exciton originally occupying the lowest available conduction band (Q/Q' valley) is scattered to a high-lying conduction band, forming a bright exciton at Γ valley, which is HX. We note that the EEA process is also possible by scattering the hole originally occupying the highest available valence band to a lower valence band, forming high-lying bright excitons at Q valley (see Supplementary Figure 3b). Supplementary Table 2 provides a summary of all potential high-lying bright exciton candidates for the upconversion from given pairs of momentum-opposite low-lying excitons. Lastly, for the resonance to be achieved, the energy of the final state (high-lying bright excitons) must be approximately twice of that of the initial state (low-lying dark excitons). Therefore, the energy conservation allows us to identify possible final states in subsequent calculations.

In order to search for the possible upconversion pathways via the resonant EEA process, quantitatively accurate exciton spectra for multilayer TMDs consist of low-lying indirect dark exciton states (K'K, QK, QΓ, KΓ) and bright exciton states covering a wide range of energy (KK$_A$, KK$_B$, QQ, ΓΓ) is demanded, rendering a numerically challenging task due to numerous involved conduction and valence bands. To overcome this obstacle, we calculate the exciton spectra by solving the BSE formulated within the WTB framework[50,51], which allows for taking the integration algorithm combined with cubic spline interpolation[51] for efficient and accurate evaluation of the Coulomb kernel in the WTB-BSE. By incorporating with the first-principles calculated quasiparticle band structures under the advanced $G_0W_0$ corrections (see Method for details), we achieved to accurately simulate the calculated exciton energies of dark $X_D$ and bright excitons ($\Gamma_{ex}$) with different CoM momentum for 1L, 2L, and 4L $WSe_2$, as presented in Figure 3c–e, respectively. In order to identify available final states for involving in the resonant EEA, we depict the energy and the doubled energy of the lowest-lying dark excitons as the grey and light green bars in the figures respectively. For 1L $WSe_2$, the lowest energy QK excitons could upconvert to high-lying KK or QQ excitons, but only the high-lying KK excitons fulfil the resonance criterion. For 2L $WSe_2$, the lowest energy QK excitons upconverting to ΓΓ excitons does not obey the selection rule (The non-resonant process is denoted with a grey arrow with red X in Fig. 3d; see item 3 of Supplementary Table 2), which explains the absence of light upconversion in our experiments (See black curve in Figure 2b). As the thickness increases, QK and QΓ excitons becomes the lowest energy states due to pronounced interlayer coupling and are likely degenerate in 4L $WSe_2$ (see Supplementary note 4 for details), and a cluster of ΓΓ excitons (green solid lines in Figure 3e) lies in the vicinity of twice the energy of QΓ/QK excitons, enabling efficient channels for the resonant EEA (denoted with red arrow in Fig. 3e). In the process of resonant EEA, a QΓ exciton is upconverted into ΓΓ excitons by absorbing the energy and momentum of either a QK or a Q'Γ exciton, which possesses an opposite momentum of the QΓ exciton (see items 4 and 6 of Supplementary Table 2). We note that only a fraction of ΓΓ excitons within the double energy of dark excitons can contribute to the UPL.

**Power dependence of light upconversion spectrum**

Our power-dependent measurement further supports the mechanism of EEA-assisted light upconversion in few-layer $WSe_2$. UPL spectra in 5L $WSe_2$ show that the overall intensity increases nonlinearly obtained with power density from 0.03 to 1.02 mW/μm$^2$ (see Supplementary Figure 7a). We integrate the intensity of UPL (including $X_{up}$ and HX) and plot it as a function of excitation power density in Figure 3f. Notably, the 5L $WSe_2$ sample exhibits a superlinear dependence with exponent α = 2 below $P$ = 0.1 mW/μm$^2$, suggesting an EEA behaviour. We note that the superlinear dependence of EEA can also be observed in other few-layer $WSe_2$ (see Supplementary Figure 7c–h). In contrast, bulk $WSe_2$ exhibits the power law exponent of α = 3, suggesting the conventional Auger recombination (see the corresponding UPL spectra in Supplementary Figure 7b). Regardless of the layer thickness, we observed a saturation of UPL intensity approaching linear dependence at $P$ > 0.1 mW/μm$^2$. The saturation may imply other higher-order nonlinear processes at elevating exciton density, which become significant at the specific power threshold.

**Resonant excitation upconversion photoluminescence spectroscopy**

We utilised a tuneable pulsed laser to selectively excite near resonantly with $X_D$ at excitation energies ($E_{in}$) from 1.24 to 1.33 eV. Figure 4a shows the evolution of UPL spectra in a 5L WSe$_2$ with various $E_{in}$. 5L WSe$_2$ has no inversion symmetry, allowing us to compare the resonant EEA and SHG. Across all the excitation energies, we observed an emission peak (yellow) strongly correlating with the $E_{in}$. Emission energies in the spectra are fitted with two Gaussians (see Methods), as displayed in Figure 4b. The extracted peak energies show a good agreement with doubled excitation energy, $2 \times E_{in}$ (depicted as the grey solid line in Figure 4b), indicating a SHG characteristic. Remarkably, when the excitation energy was tuned closely to the energy of $X_D$ peak (1.30–1.32 eV), an additional emission peak (blue) appears at 2.57 eV, which we identified as UPL, is nearly independent of $E_{in}$ (denoted as $X_{up}$ in Figure 4b). This independence is distinct from the SHG peaks because the UPL is only present at $E_{in}$ near the dark exciton energy, whereas SHG signal is observed throughout the excitation energies. We further depict normalised UPL intensity as a function of the excitation energy in Figure 4c. The intensity is significant at 1.28 eV but negligible at both 1.24 and 1.33 eV, indicating the correlation between the UPL intensity and the population of dark excitons.

**Polarisation and temperature dependence of light upconversion spectrum**

To further isolate the resonant EEA from SHG, we performed polarisation characteristics of UPL spectrum. Figure 4d shows the UPL peaks in both HH (co-polarisation) and HV (cross-polarisation) configurations[52] (see Methods for more detail) at $T$ = 80 K. The spectrum was measured with excitation energy of 2.33 eV using a CW laser. The UPL intensity is identical in different polarisation configurations, which is in contrast to the SHG where the inherently polarisation-dependent process is described by the tensor of second-order nonlinear optical susceptibility.[53] Phonon-assisted upconversion is also a mechanism through which extra energy is provided to excitons through exciton-phonon interactions, resulting in a transition to higher excitonic states.[34,35] To gain more insight into possible mechanism, we carried out temperature-dependent measurement of UPL spectrum. Figure 4e shows UPL intensity of 5L WSe$_2$ sample as a function of temperature ($T$ = 80–300 K). The UPL remains detectable up to room temperature while the intensity exponentially decreases as the temperature increases. Moreover, the energy difference between UPL energy and excitation energy is significantly higher than the energy of optical phonons in 5L WSe$_2$, which is about 50 meV (see Supplementary

Figure 8). This indicates that either exciton-phonon coupling or multi-phonon upconversion is unlikely to be responsible for the light upconversion mechanism.

**Light upconversion in few-layer transition metal dichalcogenides**

The light upconversion is not exclusive to few-layer WSe$_2$, but seen in few-layer WS$_2$, MoS$_2$, and MoSe$_2$. Figure 5a shows that the UPL emission (2.87 eV for 5L WS$_2$, 2.77 eV for 3L MoS$_2$, and 2.49 eV for 3L MoSe$_2$) closely matches the double energy of the dark exciton emissions (1.43 eV for 5L WS$_2$, 1.4 eV for 3L MoS$_2$, and 1.28 eV for 3L MoSe$_2$), signifying the important role of dark excitons in the light upconversion. Figure 5b shows that UPL intensity is strongly enhanced as the layer thickness increases, suggesting a shared characteristic of the resonant EEA process within the few-layer (Mo, W)(S, Se)$_2$. We note that the X$_D$, UPL emission, and summarised peak energy in few-layer WS$_2$ can be found in Supplementary Figure 9. Moreover, we highlight that the UPL emission is tuneable, offering a significant advantage for applications requiring specific wavelengths. Figure 5c summarises the peak energy of UPL in few-layer WSe$_2$, WS$_2$, MoS$_2$, and MoSe$_2$. The UPL emission energies span a broad spectrum, ranging from green light (2.34 eV) to ultraviolet light (3.1 eV). Such tunability illustrates the potential of these materials in enhancing optoelectronic and photocatalytic applications through engineered exciton energy, population, and dynamics.

**DISCUSSION**

The UPL via EEA of dark excitons in few-layer WSe$_2$ can be described as a two-step process: (1) a pair of dark excitons with opposite momenta converting into a high-energy exciton with zero CoM momentum and (2) the subsequent radiative recombination from high-lying bright excitons. Such EEA process is particularly efficient only when the high-lying bright excitonic states resonate with the cluster of ΓΓ excitons (see Supplementary note 3). Also, the upconverted excitons are bright—having zero CoM momentum—which enables radiative recombination. This contrasts with previously reported non-radiative EEA process; therefore, we specify the mechanism of UPL as *resonant* EEA. As for the resonant EEA above 2L WSe$_2$, the UPL emission relies on the energy overlapping between X$_{up}$ and HX, which is strongly influenced by interlayer coupling. The interlayer coupling hybridises/splits the quasiparticle band structures in few-layer TMDs. Specifically, in WSe$_2$, the splitting occurs most notably around the Γ point in the valence band largely due to the coupling of $d_{z^2}$ and $p_z$ orbitals. In contrast, the interlayer coupling at the K valley and the conduction bands at the Γ point is relatively weak, primarily involving $d_{xy}+d_{x^2-y^2}$ and $p_x+p_y$ orbitals respectively.[54,55] As a result, the interlayer coupling elevates the valence band edge at the Γ valley and lowers the exciton energies of both the indirect QΓ and the high-lying ΓΓ states, as illustrated in Supplementary Figure 4 In addition, UPL intensifies with the increasing layer thickness notably due to a high HX concentration within a narrow energy window (about 0.4 eV, see Supplementary Figures 4d,h) in thicker WSe$_2$, where the impact of the interlayer coupling on Γ-point conduction bands is relatively small, and a higher $X_D$ density in few-layer WSe$_2$ compared to bright exciton density in monolayer WSe$_2$. On one hand, the clustered ΓΓ exciton states open multiple resonant EEA channels for the upconversion from dark excitons, facilitating pronounced UPL emission as observed in our experiments. On the other hand, dipole-dipole interactions are indeed weaker in few-layer WSe$_2$, consistent with our calculation of exciton binding energy in Supplementary note 8. However, the enhanced upconversion in few-layer WSe$_2$ is related to the long lifetime of dark excitons, given the amount of $X_{up}$ is proportional to the square of $X_D$ density. This can explain upconversion efficiency $Q_{up} > 1$ in thicker WSe$_2$, which may seem to be counterintuitive.

We draw significant paragraphs on attributing upconversion mechanism to the resonant EEA. In general, energy transfer process in 2D semiconductors can be categorised as multiple photon absorption, Auger recombination, and phonon absorption.[10] First, we exclude SHG because UPL has no polarisation dependence (Figure 4d), which cannot be described by the tensor of second-order nonlinear optical susceptibility. In addition, UPL is persistently present for WSe$_2$ thickness beyond 3 layers without any evident layer dependence, as we already demonstrated in Figure 2. This is in stark contrast to SHG, which only present in odd-layer samples due to the broken inversion symmetry. It is worth noting that SHG often necessitates a femtosecond pulsed laser to achieve high photon density. Our experiments, however, have demonstrated the pronounced UPL signal simply using a CW laser, suggesting a low threshold of photon density in our upconversion process. Second, we rule out exciton-phonon scattering or phonon absorption since the decrease in UPL intensity at elevating temperatures (Figure 4e) cannot be reconciled with the increasing phonon population. Notably, the $X_{up}$ emission suggests that the radiative recombination is favoured over phonon scattering as the relaxation pathway, which typically occurs on much shorter timescales of 10–100 fs[56–58], indicating that the spontaneous emission holds a competitive edge in the dynamics of upconverted excitons in few-layer WSe$_2$. Importantly, we identify EEA as the mechanism for light upconversion in few-layer WSe$_2$ because of the quadratic dependence across 3–11L WSe$_2$ (Figure 3f and Supplementary Figure 7). The resonant EEA of dark excitons is inherently a low-dimensional phenomenon—the dominance of tightly bound excitons in few-layer WSe$_2$ due to the quantum confinement and the reduced dielectric screening, resulting in much higher exciton binding energies compared to the bulk counterpart. This hence explains the cubic dependence (Auger recombination) observed in bulk WSe$_2$, where optically excited charge carriers are unbounded or weakly bounded due to strong dielectric screening. Lastly, we note that due to the complexity of resonant EEA processes and the potential contribution of other mechanisms, such as Förster resonance energy transfer[59,60], the intricate nature of dark exciton dynamics and their associated processes requires extensive further study to be fully understood.

The resonant EEA resulting UPL emission has opened questions for future studies. Defect-related non-radiative recombination has been a competing mechanism considering the susceptibility of excitonic processes to sample imperfections. In our study, we can claim that such non-radiative recombination has a negligible impact on the upconversion across a wide range of defect density (see Supplementary note 9) within the temperature range of 80–300 K. Further investigation beyond our experimental condition, e.g., $T < 80$ K, could explore potential effect of defect-related process on the resonant EEA mechanism. From a practical perspective, a high quantum efficiency indicates a tremendous potential for optoelectronic and photocatalysis applications. To the best of our knowledge, however, it is difficult to accurately estimate UPL emission efficiency mainly because dark excitons are optically inactive. Unlike KK$_A$ bright excitons in monolayer TMDs, the lowest-lying dark exciton state of few-layer TMDs is populated through thermal relaxation of photoexcited excitons due to their indirect bandgap, making CW laser experiment hard to directly excite dark excitons. Indeed, our pulsed laser experiments directly excited dark excitons by selecting the pump energy around dark exciton energy. However, the fact that dark excitons with a finite momentum excited by a pulsed laser, which has zero momentum, still implies a contribution of phonon-exciton coupling. Therefore, a method to effectively excite dark exciton and quantify the dark exciton population in few-layer TMDs could be informative for relevant studies in the future. In our experiments, dark excitons in few-layer TMDs are generated from bright excitons through multiple phonon scatterings, a process that disrupts coherence. This intrigues potential effects of exciton coherence on EEA-assisted UPL. It could be feasible to generate a coherent pair of momentum-indirect dark excitons with opposite momenta either through the reverse process of resonant EEA or by directly exciting dark excitons using a near-field electric field.[61] This approach could enable future exploration of the role of exciton coherence in resonant EEA, potentially offering deeper insights into the dynamics and mechanisms of many-body interactions in TMDs. Aside from static studies, dynamical studies are expected to disentangle any competing processes from the resonant EEA process. Specifically, the interlayer coupling effect on the dark exciton dynamics could be analysed by pump-probe experiments. Moreover, a time-resolved ultrafast spectroscopy allows us to elucidate the dynamics of X$_{up}$ emission.

In summary, our research has showcased efficient light upconversion in atomically thin TMDs. The UPL is driven by the resonant EEA of momentum-indirect dark excitons. The demonstrated many-body interaction provides a universal route for harnessing the energy of the lowest-lying momentum-indirect excitons to generate green to ultraviolet light, featuring a broad scope of tunability. The process of exciton-mediated light conversion underlines the intricate many-body excitonic physics within TMDs and opens a door to promising advancements in the fields of optoelectronics and photocatalysis, where control over such fundamental interactions is essential.

## METHODS

### Samples preparation

Bulk WSe$_2$ and MoSe$_2$ crystals with ultra-low defects were grown by a two-step self-flux method.[62] Bulk WS$_2$ and MoS$_2$ crystals were purchased from HQ Graphene and SPI supplies, respectively. Thin TMD flakes were mechanically exfoliated onto polydimethylsiloxane (PDMS) and subsequently dry-transferred onto Al$_2$O$_3$ or SiO$_2$/Si substrates. The layer thickness of WSe$_2$ samples was determined by analysing optical contrast and ultra-low frequency Raman spectroscopy, as shown in Supplementary Figure 13.

### Upconversion Photoluminescence Spectroscopy

The fresh-made samples are transferred to a cryostat with optical access (Janis ST-500) under a high-vacuum environment (base pressure $< 5 \times 10^{-6}$ Torr) to ensure reliable results. The sample can be cooled down to 80 K by continuous flowing liquid nitrogen. A diode-pump solid-state continuous-wave laser at an excitation energy of 2.33 and 1.96 eV was employed for excitation. The spectroscopy was set up in a back-scattering configuration. A 40× objective lens (numerical aperture, 0.6) was applied to focus the laser to the diffraction limit. The PL signal was filtered by three cascaded Bragg notch filters (OptiGrate), dispersed by a grating-based monochromator (Horiba, iHR-550 with 150 gr/mm grating), and detected by a nitrogen-cooled charge-coupled device (Horiba, Symphony II). In addition, two linear polarisers are placed in incident and scattered light, which enable us to select polarisation of incident/scattered light and the relative polarisation between them. In Fig. 4d, the 0° and 90° of relative polarisation is defined as parallel (HH) and perpendicular (HV) mode, respectively. Note that all experimental figures are executed at a temperature of 80 K and incident power of 0.34 mW/μm$^2$ unless stated.

### Ultrafast Photoluminescence Spectroscopy

The ultrafast photoluminescence spectroscopy with tuneable fs pulses was executed with samples at $T$ = 9 K in a closed-cycle cryostat (Montana Instruments, Cryostation). A 20× objective lens (numerical aperture, 0.38) was used both to focus the excitation laser and collect the PL emission. The tuneable excitation came from a non-collinear optical parametric amplifier (Light Conversion, Orpheus-N-3H) pumped by the third harmonic of a Nd:YAG amplified femtosecond laser system (Light Conversion, Pharos). The spectrum of each excitation pulse was measured before the sample by a Thorlabs CCS200 spectrometer. The PL emission was separated from the pump by a 600 nm short-pass dichroic beam splitter, and measured using an Andor Kymera 328i spectrometer with an Andor Zyla sCMOS camera.

The recorded spectra were processed using a custom written python script. Each peak was fit by two Gaussian functions: one for the PL emission component and the other for the SHG component. The fits were completed using the SciPy Curve Fit module. The peak energy and full-width at half-maximum (FWHM) of the SHG component were fixed, based on the parameters from a Gaussian fit of the excitation spectrum. Specifically, the peak energy was determined by doubling the energy and the FWHM was scaled by a factor of √2 from the values obtained for the excitation spectrum. The integrated peak amplitude of the PL emission was determined from the Gaussian fit parameters.

### Normalisation of photoluminescence spectra and calculation of upconversion efficiency ($Q_{up}$)

The PL spectra of 2–6L WSe$_2$ in Figure 2a,b are normalised to 3L WSe$_2$ and the calculation of upconversion efficiency ($Q_{up}$) is based on the normalisation. Specifically, in Figure 2a, we first calculated the integrated PL intensity of dark excitons for each layer thickness. The intensity ratio of each layer thickness to 3L WSe$_2$ can be calculated using $I_{XD,n}/I_{XD,3}$, where $I_{XD,n}$ is the integrated PL intensity of dark excitons in $n$-layer WSe$_2$ ($n$ = 2–6). The spectrum of 2–6L WSe$_2$ is then scaled according to the intensity ratio. In Figure 2b, the same intensity ratio is applied to the UPL spectrum of 2–6L WSe$_2$. After the normalisation, we define the $Q_{up}$ by counting the intensity ratio of UPL to the dark exciton PL, $(I_{Xup}+I_{HX})/I_{XD}$, where $I_{Xup}$, $I_{HX}$, and $I_{XD}$ are integrated PL intensities of upconverted, high-lying, and dark excitons, respectively.

**First-principles DFT+G$_0$W$_0$ calculations of quasiparticle band structure**

The quasiparticle band structures of few-layer TMDs are obtained from the first principles calculations. Within the density functional theory (DFT) scheme, the electronic structures of few-layer WSe$_2$ were investigated through first-principles calculations. The Quantum Espresso code[63], employing norm-conserving pseudopotentials, facilitated these computations. The Perdew-Burke-Ernzerhof (PBE) functional within the generalised gradient approximation[64], supplemented with a van der Waals (DFT-D3) correction[65], was chosen to approximate the exchange-correlation interaction in the quasi-2D anisotropic systems. A plane-wave basis expansion of the wavefunction was achieved with an energy cutoff set at 80 $R$y. The Brillouin zone integration of the lattice was conducted using a 12×12×1 k-point grid sampling. Optimised lattice constants of 3.334 Å, 3.351 Å, and 3.341 Å were employed to model 1L, 2L, and 4L WSe$_2$ structures, respectively. The optimised interlayer distance of 2L and 4L WSe$_2$ are $d^{2L}$ = 3.333 Å and $d^{4L}$ = 3.342 Å, respectively. Furthermore, to address accurate many-body effects beyond mean-field theory, advanced quasiparticle (QP) G$_0$W$_0$ calculations were performed using the BerkeleyGW package[66], based on the DFT results. A nonuniform neck subsampling (NNS) scheme[5] was employed to ensure efficient convergence with q-point sampling. The QP gaps were converged to better than 0.1 eV utilising a dielectric cutoff of 35 $R$y and up to 2600 bands.

**WTB-BSE calculations of exciton spectra of multilayer TMDs**

The exciton state can be expressed in terms of the superposition of electron-hole configurations as $|X_{S\mathbf{Q}}\rangle = \frac{1}{\sqrt{\mathcal{A}}}\sum_{vc\mathbf{k}} \hat{c}^\dagger_{c\mathbf{k}+\mathbf{Q}}\hat{c}_{v\mathbf{k}}|GS\rangle$, where $\mathcal{A}$ is the area of the two-dimensional material, $S$ and $\mathbf{Q}$ label the quantum number and wave vector of the exciton state, respectively, with $\hat{c}^{(\dagger)}_{c/v\mathbf{k}}$ being the annihilation (creation) operator of the electron in conduction/valence band (CB/VB) $c/v$ carrying momentum $\mathbf{k}$. The exciton energy $E^X_{S\mathbf{Q}}$ and the wavefunction $A^{X,S\mathbf{Q}}_{vc\mathbf{k}}$, where the latter characterises the amplitude of electron-hole configurations $\hat{c}^\dagger_{c\mathbf{k}+\mathbf{Q}}\hat{c}_{v\mathbf{k}}|GS\rangle$, are determined by the Bethe-Salpeter equation (BSE)

$$(\epsilon_{c\mathbf{k}+\mathbf{Q}} - \epsilon_{v\mathbf{k}})A^{X,S\mathbf{Q}}_{vc\mathbf{k}} - \frac{1}{\mathcal{A}}\sum_{v'c'\mathbf{k}'}\left[W^{eh,d}_{c\mathbf{k}+\mathbf{Q},v'\mathbf{k}',v\mathbf{k},c'\mathbf{k}'+\mathbf{Q}} - V^{eh,x}_{c\mathbf{k}+\mathbf{Q},v'\mathbf{k}',c'\mathbf{k}'+\mathbf{Q},v\mathbf{k}}\right]A^{X,S\mathbf{Q}}_{v'c'\mathbf{k}'} = E^X_{S\mathbf{Q}}A^{X,S\mathbf{Q}}_{vc\mathbf{k}}. \quad (1)$$

Here, $\epsilon_{c/v\mathbf{k}}$ is the quasiparticle energy of conduction/valence band $c/v$, $W$ and $V$, with the superscripts respectively indicating the electron-hole direct and exchange interactions, are the Coulomb matrix elements defined as

$$W_{c\mathbf{k}+\mathbf{Q},v'\mathbf{k}',v\mathbf{k},c'\mathbf{k}'+\mathbf{Q}} = \int d^3\mathbf{r}_1 d^3\mathbf{r}_2\, \psi^*_{c\mathbf{k}+\mathbf{Q}}(\mathbf{r}_1)\psi^*_{v'\mathbf{k}'}(\mathbf{r}_2)W(\mathbf{r}_1,\mathbf{r}_2)\psi_{v\mathbf{k}}(\mathbf{r}_2)\psi_{c'\mathbf{k}'+\mathbf{Q}}(\mathbf{r}_1)\,, \quad (2)$$

$$V_{c\mathbf{k}+\mathbf{Q},v'\mathbf{k}',c'\mathbf{k}'+\mathbf{Q},v\mathbf{k}} = \int d^3\mathbf{r}_1 d^3\mathbf{r}_2\, \psi^*_{c\mathbf{k}+\mathbf{Q}}(\mathbf{r}_1)\psi^*_{v'\mathbf{k}'}(\mathbf{r}_2)V(\mathbf{r}_1-\mathbf{r}_2)\psi_{c'\mathbf{k}'+\mathbf{Q}}(\mathbf{r}_2)\psi_{v\mathbf{k}}(\mathbf{r}_1)\,, \quad (3)$$

where, $\psi_{c/v\mathbf{k}}(\mathbf{r})$ is the single-electron wavefunction, $V(\mathbf{r}) = \frac{e^2}{4\pi\varepsilon_0|\mathbf{r}|}$ and $W(\mathbf{r}_1, \mathbf{r}_2) = \int d^3\mathbf{r}'\varepsilon^{-1}(\mathbf{r}_1, \mathbf{r}')V(\mathbf{r}' - \mathbf{r}_2)$ are the bare and screened Coulomb potentials, respectively, and $\varepsilon^{-1}(\mathbf{r}_1, \mathbf{r}_2)$ is the inverse dielectric function.

In the computations of the BSE[51], the electron-hole Coulomb matrix elements are formulated in the Wannier tight-binding (WTB) framework which unitarily connects the quasiparticle Bloch wave functions with the maximally localised Wannier functions (WFs) $\mathcal{W}_{j\mathbf{R}}(\mathbf{r})$ by

$$\psi_{n\mathbf{k}}(\mathbf{r}) = \frac{1}{\sqrt{N}}\sum_j C_j^{(n)}(\mathbf{k}) \sum_\mathbf{R} e^{i\mathbf{k}\cdot\mathbf{R}}\mathcal{W}_{j\mathbf{R}}(\mathbf{r}), \tag{4}$$

where $j$ labels the orbital-like WFs, $\mathbf{R}$ indicates the unit cell that the WF locates at, and the coefficients $C_j^{(n)}(\mathbf{k})$ are determined by the WTB Hamiltonian $H_{j\ell}^{\mathrm{WTB}}(\mathbf{k}) = \sum_\mathbf{R} e^{i\mathbf{k}\cdot\mathbf{R}}\langle\mathcal{W}_{j\mathbf{0}}|\widehat{H}^{\mathrm{KS}}|\mathcal{W}_{\ell\mathbf{R}}\rangle$, which is equivalent to the (G$_0$W$_0$-corrected) Kohn-Sham Hamiltonian $\widehat{H}^{KS}$ represented in the WF basis[50]. Within the WTB framework, the Coulomb kernels (2, 3) of the BSE are computed efficiently and accurately by utilising the algorithms based on analytic integrations and local cubic spline interpolation[51].

For the purpose of searching the candidates of HXs for the resonant EEA, the electron-hole exchange interaction is neglected since it is a minor factor. In constructing the electron-hole direct Coulomb matrix elements, the screened Coulomb potential is obtained by solving the Poisson's equation in $\mathbf{k}$-space for layered structures[67]. Here, we take the effective thickness of 6.72 Å, 11.70 Å, 23.75 Å for 1L, 2L, 4L WSe$_2$, respectively, and the dielectric constant of the environment is set to $\varepsilon = 1$ (air). The resulting exciton spectra for 1L, 2L, and 4L WSe$_2$ are presented in Figure 3c–e in the main text. Note that, for clarity, we do not show the high-lying excitonic Rydberg series (e.g. 2s, 2p) which are less relevant to the upconversion photoluminescence.

## DATA AVAILABILITY

The data generated for this study have been deposited in the Figshare database.

## CODE AVAILABILITY

The first-principles and the G$_0$W$_0$ calculations were performed by using the Quantum ESPRESSO (https://www.quantum-espresso.org/) and BerkeleyGW (https://berkeleygw.org/), respectively. Wannierization processes were performed using Wannier90 (https://wannier.org/). The in-house codes of the WTB-BSE (https://quantum.web.nycu.edu.tw/wannierbse/) being under developed toward licensed open-source code are available from the corresponding author upon request.

**ACKNOWLEDGMENT**

Y. H. C., K.W. B., J. A. D., M. S. F., and S. Y. C. acknowledge support from the ARC Centre of Excellence in Future Low-Energy Electronics Technologies (FLEET; CE170100039). S. Y. C. acknowledges support from the National Science and Technology Council of Taiwan through Grant 111-2112-M-002-047, 112-2628-M-002-008-, and 113-2628-M-002-020, and the Center of Atomic Initiative for New Materials, National Taiwan University (grant nos. 112 L9008 and 113 L9008), from the Featured Areas Research Center Program within the framework of the Higher Education Sprout Project by the Ministry of Education of Taiwan. Y. H. C. and S. Y. C. thank the experimental support from Dr. Wei-Hua Wang in the Institute of Atomic and Molecular Sciences, Academia Sinica. S. J. C. acknowledges that this study is supported by the National Science and Technology Council of Taiwan under the contract, 112-2112-M-A49-028- , and by National Center for High-Performance Computing of Taiwan. P Y. L. acknowledges support from the National Science and Technology Council of Taiwan, under the contract NSTC 112-2112-M-A49-019-MY3. H. C. H., C. E. H., and Y. N. H. acknowledge the support from National Science and Technology Council, Taiwan, under Grant: 110-2112-M-032-014-MY3, 113-2112-M-032-013, and thank the National Center for High-Performance Computing in Taiwan for providing computational resources. Tungsten diselenide and molybdenum diselenide crystal growth was supported under the United States National Science Foundation Materials Research Science and Engineering Center through grants DMR-1420634 and DMR-2011738. P. H. work was supported by the Quantum Sensors QSP078 and On-Chip Integrated Photonic Circuits Based on 2D Materials HTSN341 Challenge Programs at the National Research Council of Canada, NSERC Discovery Grant No. RGPIN-2019-05714, and University of Ottawa Research Chair in Quantum Theory of Materials, Nanostructures, and Devices.

**Author Contributions**

Y. H. C. and P. Y. L. contributed equally to this work. Y. H. C. and S. Y. C. conceived and designed this project. Y. H. C., S. Y. C., and M. S. F. conducted device fabrication, optical measurement, and data analysis. J. H., L. N. H., and K. B. from Columbia University provided bulk $WSe_2$ and $MoSe_2$ crystals. M. H. contributed towards confirming the quality of TMD crystals. K. B. and J. A. D. from Swinburne University of Technology conducted ultrafast optical measurement. P. Y. L., G. H. P., C. J. H., C. E. H., Y. N. H., P. H., H. C. H., and S. J. C. contributed theoretical calculations. W. H. W. provided resources for conducting optical measurements. The manuscript was written through contributions of all authors. All authors have given approval to the final version of the manuscript.

**Competing Interests Statement**

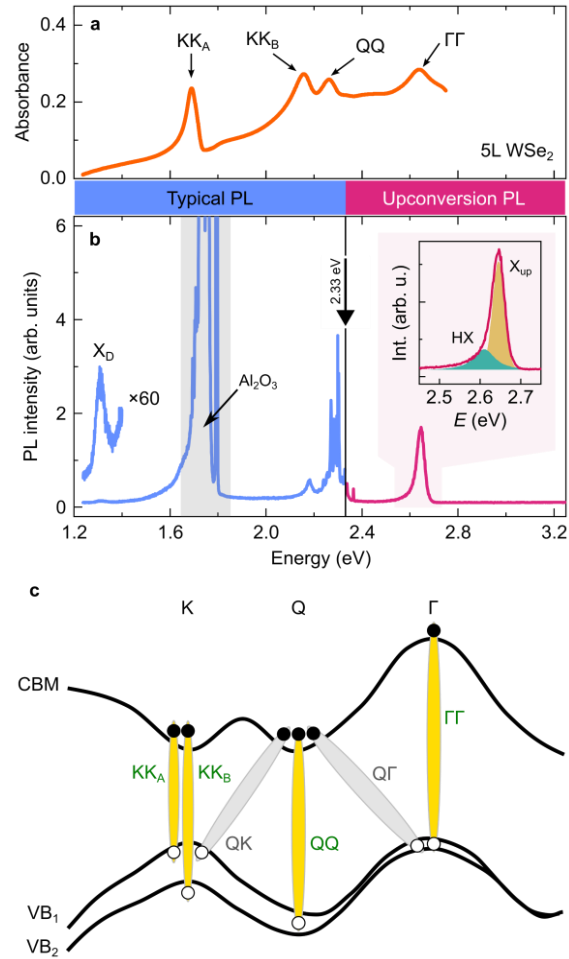

**Figure 1.** Bright, dark and upconverted excitons in 5L $WSe_2$. **a** Absorbance spectrum at $T$ = 80 K with bright excitons indicated by arrows; labels discussed in text. **b** PL spectrum at $T$ = 80 K excited with excitation energy 2.33 eV (wavelength 532 nm). The spectrum is divided into typical (blue) and upconversion PL (ruby) regimes. The inset in upconversion PL regime: The peak fittings of UPL from upconverted excitons $X_{up}$ and high-lying excitons (HX). The low-energy PL is also shown magnified 60× to better show the dark exciton $X_D$ emission at 1.31 eV. The grey shadow area indicates photoluminescence from the $Al_2O_3$ substrate. **c** The schematic of electronic band structure and corresponding bright/dark excitons observed in (**a**,**b**).

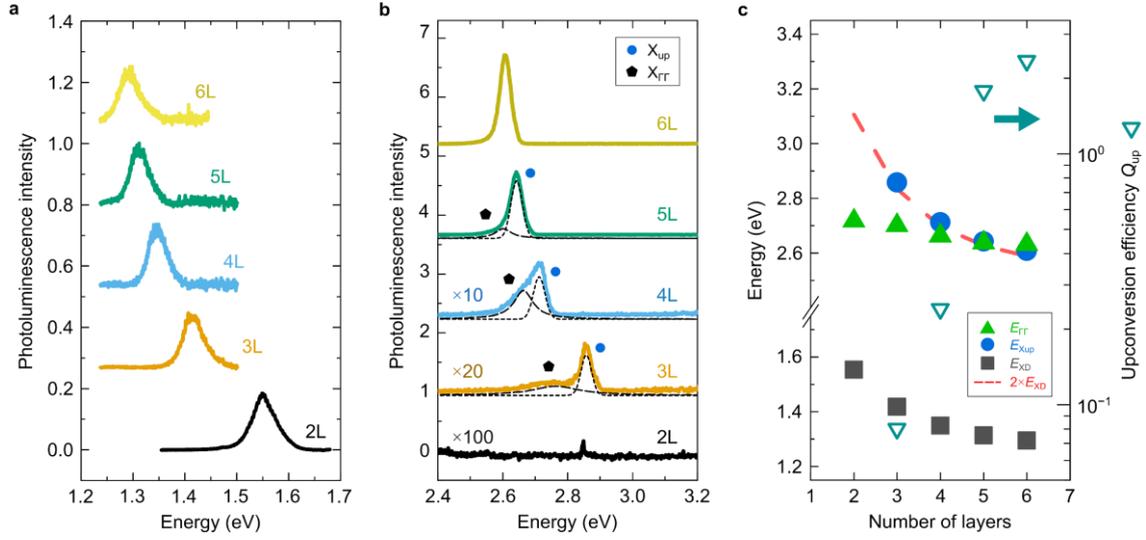

**Figure 2.** Layer-dependent characteristics of UPL in few-layer $WSe_2$ at $T = 80$ K. Evolution of **a** dark excitons $X_D$ and **b** UPL spectra with the number of layers from 2L to 6L. The UPL emission of 3–6L $WSe_2$ can be deconvoluted into upconverted excitons $X_{up}$ (short dash line; peak denoted by circles) and high-lying excitons HX (long dash line; peak denoted by pentagons). The spectrum of each thickness is vertically shifted for clarity. **c** Emission energy of $X_D$ and $X_{up}$, and absorption energy of ΓΓ excitons $E_{ΓΓ}$, and extracted upconversion efficiency $Q_{up}$ as a function of layer thickness. In (**a**–**c**), the photoluminescence intensity is normalised to $X_D$ emission of 3L $WSe_2$ (see Method for further details).

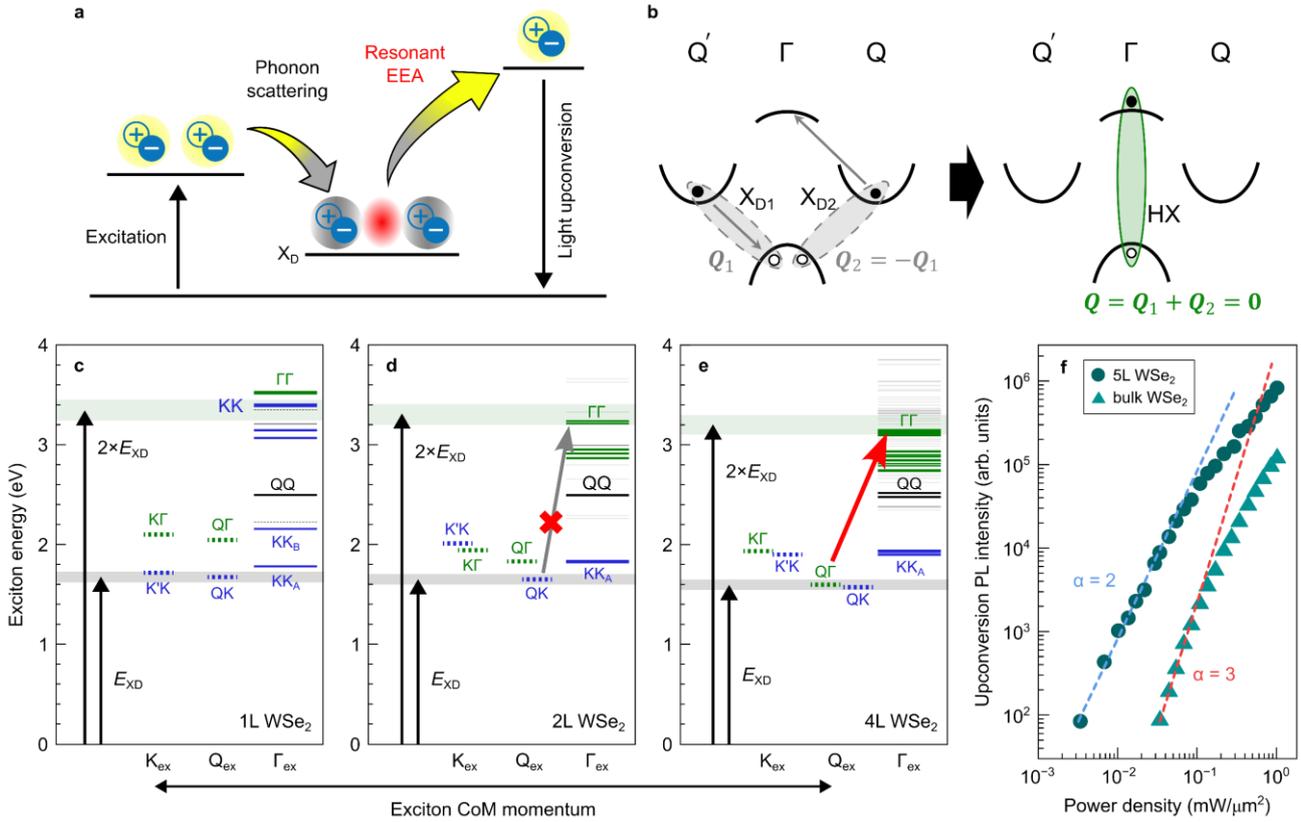

**Figure 3.** Resonant exciton-exciton annihilation of dark excitons in few-layer WSe$_2$. **a** The schematic of upconversion process in few-layer WSe$_2$. Light-excited bright excitons thermally relax to dark exciton states X$_D$ at the lowest energy through multiple phonon scattering. The resonant EEA process upconverts two dark excitons X$_D$ to a single high-energy exciton X$_{up}$, followed by a radiative combination, resulting in the light upconversion emission. **b** A schematic representation of excitonic upconversion before (left panel) and after (right panel) the electron-scattering process, which is indicated by grey arrows. **c**–**e** WTB-based BSE calculated exciton energy as a function of exciton centre of mass (CoM) momentum in 1L (**c**), 2L (**d**), and 4L WSe$_2$ (**e**). Black arrows denote the energy of the lowest energy dark exciton $E_{XD}$ and the corresponding twice energy 2×$E_{XD}$. **f** UPL intensity as a function of power density for 5L and bulk WSe$_2$ obtained at $T$ = 80 K. The blue and red dash lines represent a power-law exponent α = 2 and α = 3, respectively.

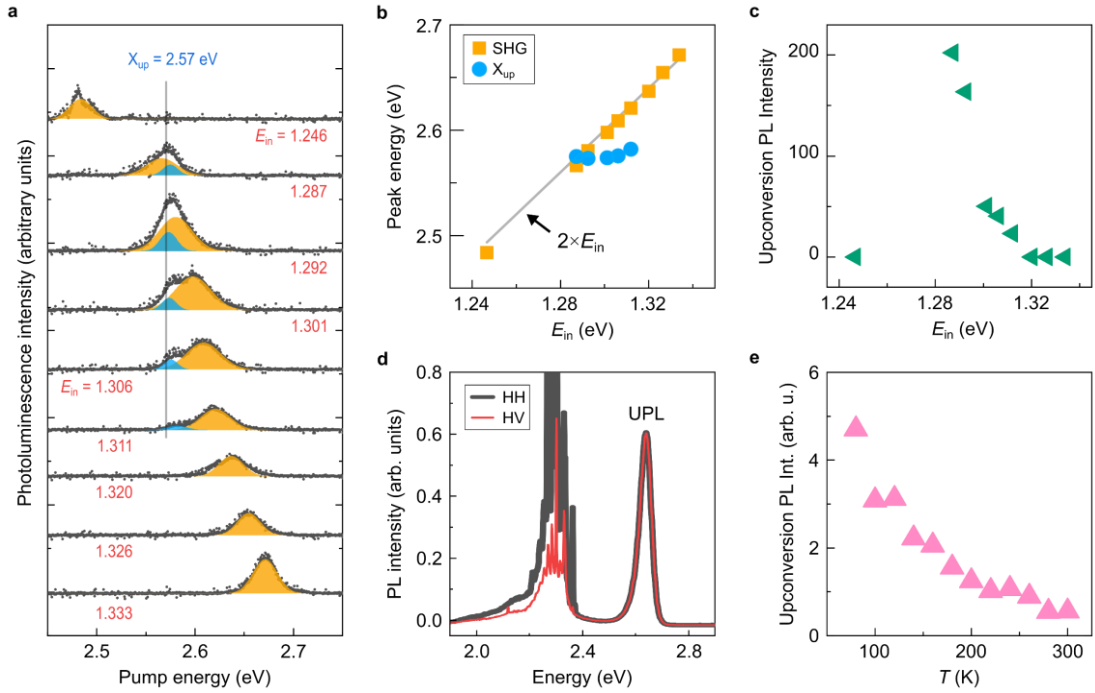

**Figure 4.** Excitation energy, polarisation, and temperature dependence of light upconversion in 5L $WSe_2$. **a** Evolution of UPL spectrum at $T$ = 9 K with excitation energy from 1.246 to 1.333 eV. The excitation energy for each spectrum is labelled accordingly. Blue and orange peaks represent UPL emission from upconverted excitons $X_{up}$ at 2.575 eV and SHG, respectively. **b** Extracted energy of $X_{up}$ and SHG as a function of excitation energy. The grey solid line indicates the double of the excitation energy. **c** Normalised UPL intensity as a function of excitation energy. The intensity is normalised by the square of incident power. **d** UPL spectra at $T$ = 80 K obtained in perpendicular (HV) and parallel (HH) scattering configuration. **e** UPL intensity as a function of temperature ranging from 80 to 300 K. In (**a**–**c**), the spectra were performed using the ultrafast PL spectroscopy (see Method); the spectra in (**d**,**e**) were obtained using a continuous wave laser with an excitation energy of 2.33 eV.

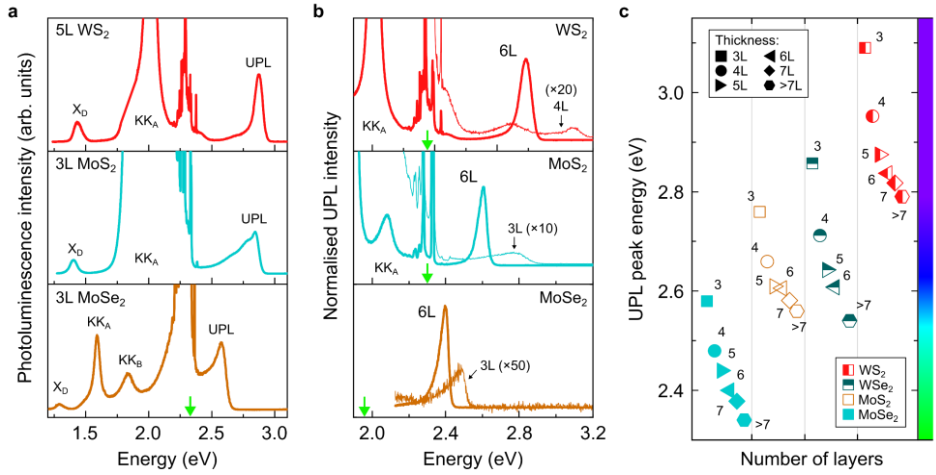

**Figure 5.** Light upconversion in few-layer MoS$_2$, MoSe$_2$, WS$_2$, and WSe$_2$ at $T$ = 80 K. **a** Photoluminescence spectrum of 5L WS$_2$, 3L MoS$_2$, and 3L MoSe$_2$ at emission energy ranging from 1.2 to 3.1 eV. The excitation energy (2.33 eV) is denoted as the green arrow. **b** UPL spectra of 6L WS$_2$, MoS$_2$, and MoSe$_2$ in contrast to thinner counterparts (3L WS$_2$, 3L MoS$_2$, and 4L MoSe$_2$). The spectra are normalised to the incident power. The excitation energy for each spectrum (2.33 eV for WS$_2$ and MoS$_2$, 1.96 eV for MoSe$_2$) is denoted as the green arrow. **c** Peak energy of UPL in MoS$_2$ (open), MoSe$_2$ (solid), WS$_2$ (half left), and WSe$_2$ (half up) with the layer thickness ranging from 3L to 7L, and thicker than 7L.